\documentclass{agujournal}
\draftfalse

\journalname{Journal of Geophysical Research - Space Physics}

\usepackage{savesym}
\savesymbol{tablenum}
\usepackage[output-decimal-marker={.}, exponent-product=\cdot, per-mode=fraction]{siunitx}
\restoresymbol{SIX}{tablenum}

\usepackage{amsmath, amssymb, bm}

\usepackage{longtable}
\usepackage{url}

\usepackage{color}

\begin{document}
\title{The Dependence of the Peak Velocity of High-Speed Solar Wind Streams as Measured in the Ecliptic by ACE and the STEREO satellites on the Area and Co-Latitude of their Solar Source Coronal Holes}

\authors{Stefan J. Hofmeister\affil{1}, Astrid Veronig\affil{1}, Manuela Temmer\affil{1}, Susanne Vennerstrom\affil{2}, Bernd Heber\affil{3}, Bojan Vr\v{s}nak\affil{4}}

\affiliation{1}{University of Graz, Institute of Physics, IGAM-Kanzelh\"ohe Observatory, NAWI Graz, Graz, Austria}
\affiliation{2}{National Space Institute, DTU Space, Denmark}
\affiliation{3}{Institut f\"ur Experimentelle und Angewandte Physik, Universit\"at Kiel, Kiel, Germany}
\affiliation{4}{Hvar Observatory, Faculty of Geodesy, Zagreb, Croatia}

\correspondingauthor{Stefan J. Hofmeister}{stefan.hofmeister@uni-graz.at}


\begin{abstract}
We study the properties of 115 coronal holes in the time-range from 2010/08 to 2017/03, the peak velocities of the corresponding high-speed streams as measured in the ecliptic at \SI{1}{AU}, and the corresponding changes of the Kp index as marker of their geo-effectiveness.
We find that the peak velocities of high-speed streams depend strongly on both the areas and the co-latitudes of their solar source coronal holes with regard to the heliospheric latitude of the satellites. Therefore, the co-latitude of their source coronal hole is an important parameter for the prediction of the high-speed stream properties near the Earth.
We derive the largest solar wind peak velocities normalized to the coronal hole areas for coronal holes located near the solar equator, and that they linearly decrease with increasing latitudes of the coronal holes. For coronal holes located at latitudes $\gtrsim \SI{60}{\degree}$, they turn statistically to zero, indicating that the associated high-speed streams have a high chance to miss the Earth. 
Similar, the Kp index per coronal hole area is highest for the coronal holes located near the solar equator and strongly decreases with increasing latitudes  of the coronal holes. 
We interpret these results as an effect of the three-dimensional propagation of high-speed streams in the heliosphere, i.e., high-speed streams arising from coronal holes near the solar equator propagate in direction towards and directly hit the Earth, whereas solar wind streams arising from coronal holes at higher solar latitudes only graze or even miss the Earth.
\end{abstract}

\section{Introduction}


Since the 1970s, it is well known that solar coronal holes, i.e., coronal regions with a reduced density and temperature as compared to the ambient corona and an open magnetic field topology, are the source of high-speed solar wind streams, i.e., supersonic plasma streams transcending our solar system \citep{nolte1976}. The supersonic plasma streams propagate radially away from the rotating Sun and form a branch of the Parker's spiral \citep{parker1958}. Thereby, they compress the preceding plasma of the slow solar wind and form a shock region, known as stream interaction region (SIR). Whenever the high-speed solar wind streams and the associated SIRs hit the Earth, they compress the Earth's magnetosphere and may cause geomagnetic storms. Since high-speed solar wind streams are the major cause of minor and medium geomagnetic storms at Earth at the solar declining phase \citep{richardson2000}, and since high-speed solar wind streams are thought to precondition the interplanetary space and the state of the Earth's magnetosphere for subsequent stronger events like coronal mass ejections \citep[CMEs;][]{gonzalez1996}, the forecast for the properties of high-speed solar wind streams is of high interest.   

The current real-time forecast models for the velocity of high-speed solar wind streams near the Earth are based on simulations of the heliosphere and of the solar corona (e.g., ENLIL, \citet{odstrcil2003}), on an empirical relationship to the flux tube expansion factor of the magnetic field evaluated between the bottom of coronal holes and the source surface at about $2.5 R_\odot$ (e.g.\ the Wang-Sheeley-Arge model; \citealt{arge2000, arge2003}), and/or on a statistical relationship to the area of coronal holes (e.g. the Empirical Solar Wind Forecast; \citealt{rotter2012, reiss2016}). Note that the empirical and statistical forecast models are related to each other: \citet{fainshtein1994} showed that the area, the flux tube expansion factor, and the photospheric mean magnetic field density below the bottom of coronal holes depend on each other.

In the following, we focus on the statistical relationship between the area of coronal holes and the velocity of high-speed streams as measured in the ecliptic at \SI{1}{AU}.
In 1976, \cite{nolte1976} showed that the areas of 3 low-latitude coronal holes which crossed the central meridian in total 15 times between 1973/05 and 1974/02 correlate with the peak velocities of the corresponding high-speed solar wind streams with a Pearson correlation coefficient $cc=0.96$.
\cite{abramenko2009} extended the study to 44 single low-latitude coronal holes observed between 2001 and 2006 and found that their area correlates with the peak velocity of the corresponding high-speed streams at L1 with $cc = 0.75$. \cite{karachik2011} revealed that the projected areas of 108 single coronal holes observed between 1998 and 2008 also correlate with the peak velocity at L1 with $cc=0.41-0.65$, with the highest correlation at medium solar activity, i.e., at the rising and declining phase of the solar cycle. 
Further, \cite{wang1990} studied the three-month-averages of the total area coronal holes covering on the Sun's disk between 1967 and 1988 and found that they correlate with the three-month-averages of the velocity of high-speed streams measured near the Earth. \cite{vrsnak2007a} derived that also the area of coronal holes within a meridional slice of [-10$^\circ$, 10$^\circ]$ correlates well with the peak velocity of high-speed streams measured at L1 with $cc=0.62$, studying a period of 100 days in 2005. Both \cite{wang1990} and \cite{vrsnak2007b} reported that the correlations degrade when polar coronal holes were excluded.

Further, \cite{vrsnak2007b} showed that the total area coronal holes cover within a meridional slice of [-10$^\circ$, 10$^\circ]$ also correlates with the drop of the geomagnetic Dst index induced by the impacting high-speed solar wind streams. The Pearson correlation coefficient between the areas and the Dst index is $cc=0.31$. The correlation increases to $cc=0.86$ when taking into account the Russel-McPherron effect \citep{russel1973}. Again, the correlation decreases if the polar coronal holes are excluded from the analysis.

Note that these results are statistical relationships between the coronal hole area, the peak velocities of high-speed solar wind streams at L1, and the strength of geomagnetic storms. They neglect the three-dimensional propagation of high-speed solar wind streams in the heliosphere and thus the three-dimensional nature of the relation between coronal hole areas, high-speed solar wind stream peak velocities at L1, and strengths of geomagnetic storms.

The three-dimensional distribution of the solar wind in the inner heliosphere was first investigated by the satellite Ulysses \citep{marsden2001}. 
Investigations based on data from Ulysses showed that at solar minimum, the heliospheric distribution of the solar wind is dominated by high-speed solar wind streams from medium to high heliospheric latitudes arising from large polar coronal holes, and by the slow solar wind streams near the ecliptic \citep{mccomas2000}. In contrast, at solar maximum, both slow and high-speed solar wind streams are apparent from the ecliptic up to high latitudes \citep{mccomas2001}. Further, Ulysses sampled the heliospheric distribution of a high-speed solar wind stream arising from a stable coronal hole at solar maximum. They found that the heliospheric velocity distribution strongly depended on the boundary of the polar coronal hole,  and that the corresponding high-speed stream expanded down to $\approx 55-70 \si{\degree}$ heliospheric latitude \citep{mccomas2003b}.

While, it is not clear (1) whether the area of single coronal holes or the total area coronal holes cover on the solar disk is the better predictor for the high-speed solar wind stream peak velocity at L1, (2) how the presence of large polar coronal holes contribute to the speeds of high-speed solar wind streams arising from single low- and mid-latitude coronal holes, (3) how the morphology of coronal holes affects the speeds of high-speed streams, (4) how the relationship between high-speed stream peak velocities at L1 and coronal hole areas is affected by the three-dimensional propagation of high-speed solar wind streams in the heliosphere, and (5) how the relationship between high-speed stream peak velocities and coronal hole areas change over the solar cycle.


In this paper, we analyse the properties of 115 coronal holes observed by the satellites SDO, STEREO A, and STEREO B distributed over all latitudes between 2010 and 2017, their relationship to the peak velocity of their related high-speed solar wind streams measured by the satellites ACE, STEREO A, and STEREO B, and their relationship to the strength of geomagnetic storms induced by the high-speed solar wind streams for a subset of the 52 Earth-directed high-speed solar wind streams.
Besides the well-known relationship between the high-speed solar wind stream peak velocities and the areas of their source coronal holes, we find a distinct relationship to the co-latitude of their solar source coronal holes, and interpret this result as a consequence of the three-dimensional propagation and expansion of high-speed solar wind streams in the heliosphere.

The paper is structured as follows: Section \ref{datared} describes shortly the datasets used and the data reduction performed, Section \ref{analysis} performs the analysis. Section \ref{results} presents the results: Section \ref{hss_v} shows the dependency of the peak velocities of high-speed streams as measured in the ecliptic at \SI{1}{AU} on the area and solar latitude of their source coronal holes, and Section \ref{Kp} the dependency of the Kp index on the area and solar latitude of the source coronal holes. In Section \ref{discussion} we discuss the results.

\section{Datasets and Data Reduction} \label{datared}

To determine the properties of the coronal holes, we use EUV \SI{193}{\angstrom} filtergrams recorded by the Atmospheric Imaging Assembly (AIA) on-board the Solar Dynamics Observatory \citep[SDO;][]{lemen2012} and provided by the Joint Science Operations Center (JSOC; \url{http://jsoc.stanford.edu/}), and EUV \SI{195}{\angstrom} filtergrams recorded by the Extreme UltraViolet Imagers (EUVI) on-board of the twin-satellites STEREO A and STEREO B \citep[STA; STB;][]{howard2008} and provided by the Virtual Solar Observatory (VSO; \url{https://sdac.virtualsolar.org/cgi/search}).
The AIA \SI{193}{\angstrom} filtergrams show the emission from the Fe XII ions in the coronal plasma at a temperature of \SI{1.6}{MK} (peak response), and the EUVI \SI{195}{\angstrom} filtergrams the emission from the Fe XII ions at a temperature of \SI{1.4}{MK} (peak response).
All images were normalized to an exposure time of \SI{1}{s}, rotated to solar north, and rescaled to a spatial resolution of \SI{2.4}{arcsec/pixel} by considering the conservation of flux. 

To analyse the velocities of high-speed solar wind streams, we use in-situ solar wind bulk velocity measurements from the Solar Wind Electron, Proton, and Alpha Monitor \citep[SWEPAM;][]{mccomas1998} on-board of the Advanced Composition Explorer (ACE) and provided by Caltech (\url{http://www.srl.caltech.edu/ACE/ASC/level2/index.html}), and in-situ solar wind bulk velocity measurements from the PLasma And Supra-Thermal Ion Composition investigation \citep[PLASTIC;][]{galvin2008} instrument on-board of STA and STB and provided by the PLASTIC consortium (\url{http://aten.igpp.ucla.edu/forms/stereo/level2_plasma_and_magnetic_field.html}). The SWEPAM-Ion instrument is a spherical section electrostatic energy per charge analyser, measuring the energy of solar wind ions from \SIrange{0.26}{35}{keV}, which is dominated by solar wind protons \citep{mccomas1998}. Based on this data, Caltech provides a level 2 dataset containing the absolute values of the hourly-averaged bulk solar wind speed.
The PLASTIC instrument is an electrostatic energy per charge and time-of-flight analyser, measuring the solar wind proton bulk parameters in an energy range from \SIrange{0.3}{10.6}{keV}.

For the study of the geomagnetic consequences of the Earth-directed high-speed streams, we use the geomagnetic Kp-index. The Kp index is a measure of the disturbances of the horizontal component of the Earth's magnetic field strength averaged over 13 observatories located between \SIrange{44}{60}{\degree} latitude \citep{bartels1949, menvielle1991}. Here, we use the Kp index as listed in the OMNI database (\url{https://omniweb.gsfc.nasa.gov/form/dx1.html}).

\section{Methods} \label{analysis}

\begin{figure*}[p]
\centering
\includegraphics[width = \textwidth]{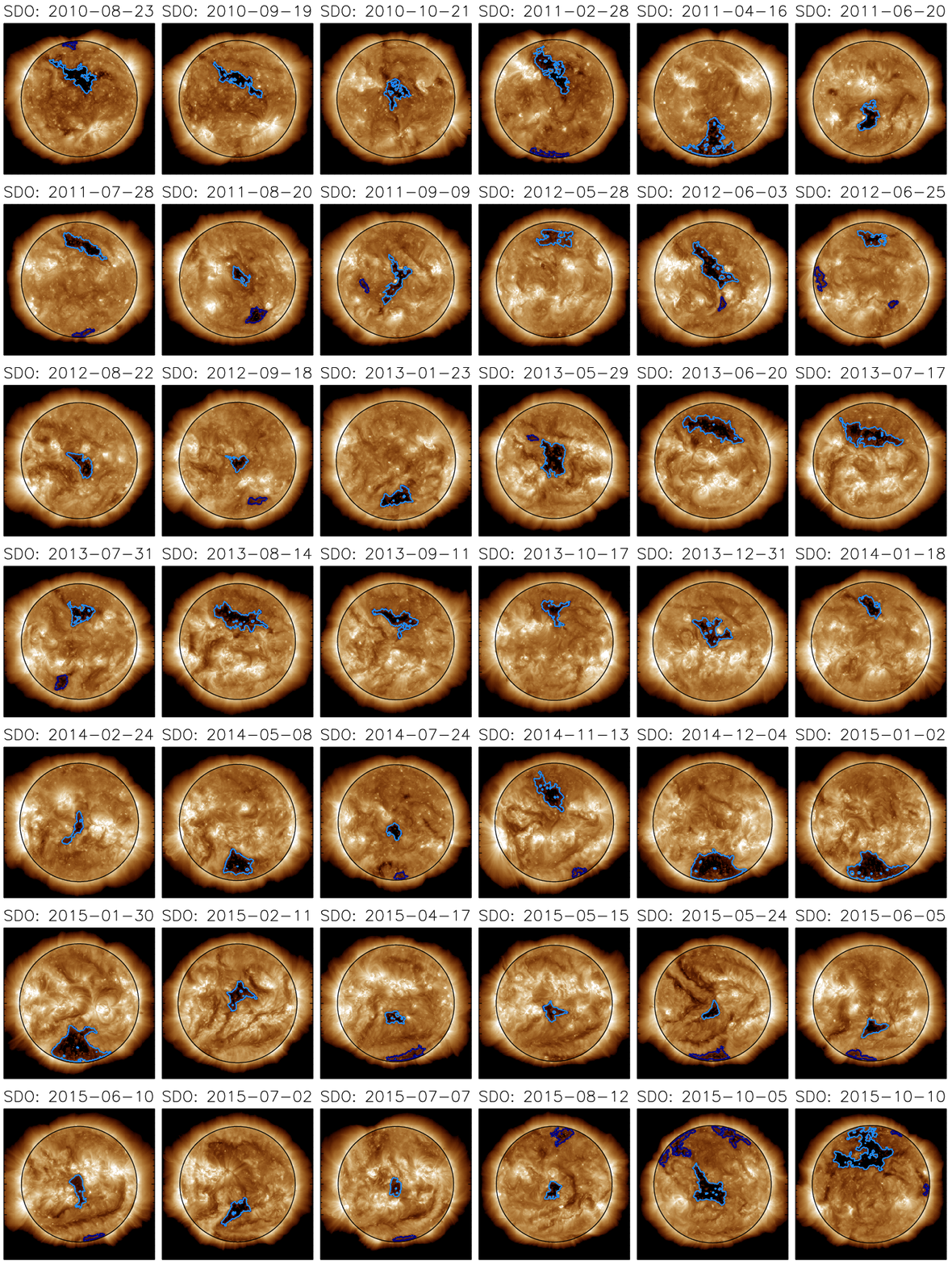}
\caption{SDO/AIA-193 filtergrams of the coronal holes. All detected coronal holes are outlined; the coronal holes used in the study are marked in light blue.}
\label{37ch_1}
\end{figure*}

\begin{figure*}[p]
\centering
\includegraphics[width = \textwidth]{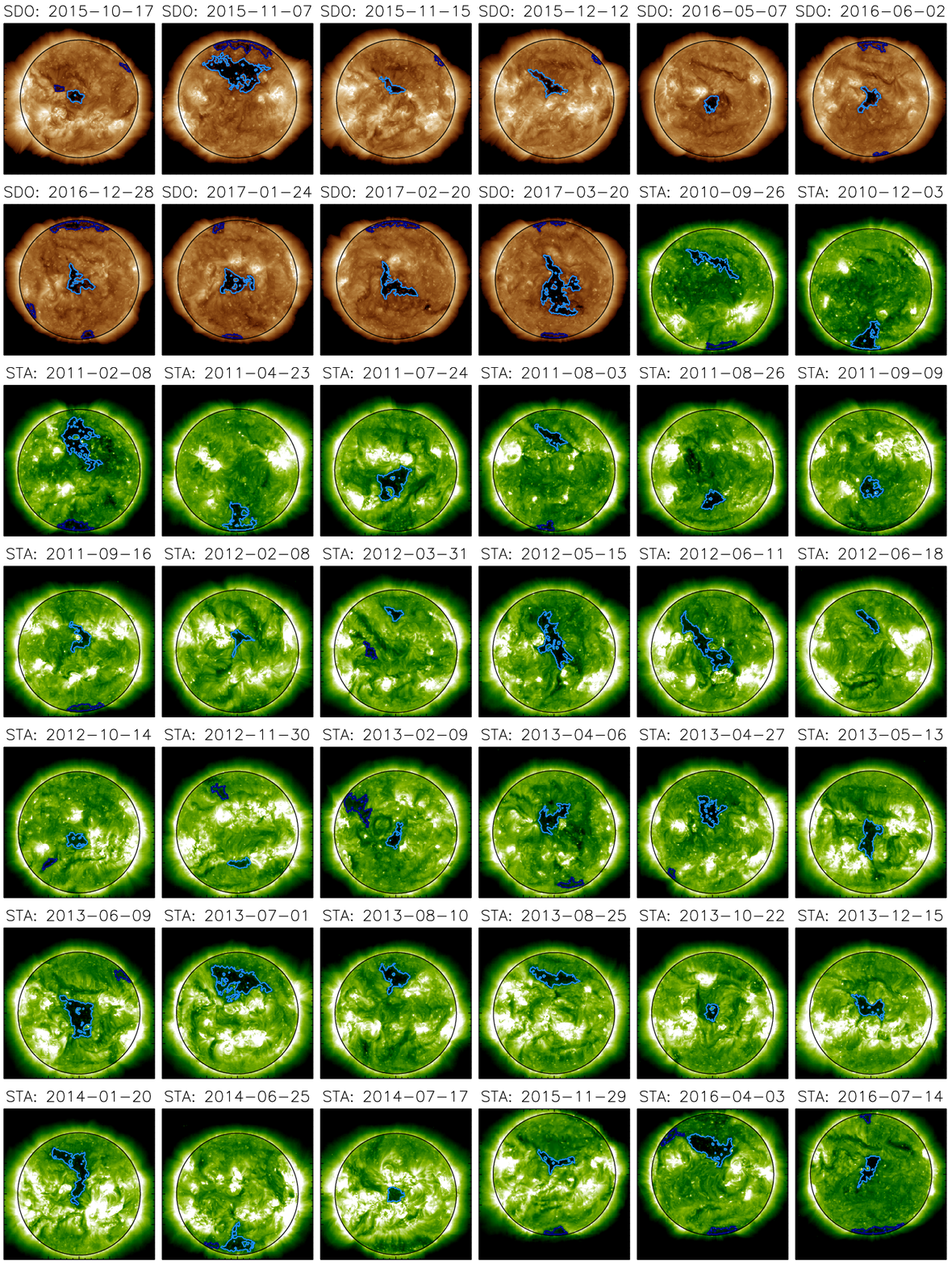}
\caption{SDO/AIA-193 and STA/EUVI-195 filtergrams of the coronal holes. All detected coronal holes are outlined; the coronal holes used in the study are marked in light blue.}
\label{37ch_2}
\end{figure*}

\begin{figure*}[p]
\centering
\includegraphics[width = \textwidth]{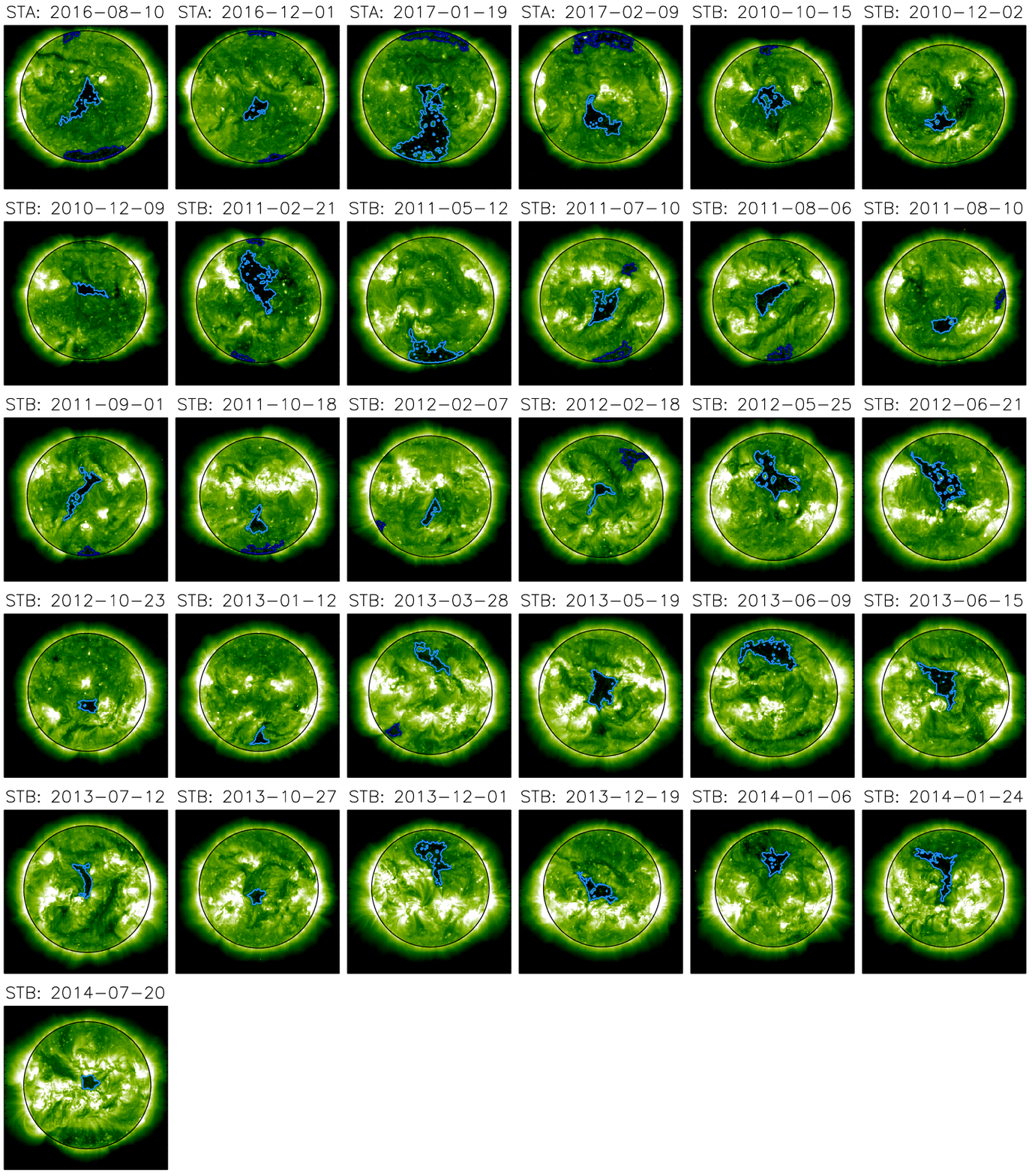}
\caption{STB/EUVI-195 filtergrams of the coronal holes. All detected coronal holes are outlined; the coronal holes used in the study are marked in light blue.}
\label{37ch_3}
\end{figure*}

We manually selected 115 solar coronal holes and the corresponding high-speed solar wind streams in the time range from 2010/08 to 2017/03. Note that we define every solar wind stream arising from a coronal hole which produces a stream interaction region as a high-speed solar wind stream.
The criteria for choosing the events was that 
\begin{itemize}
\item the coronal holes show a well-defined boundary as seen in the SDO/AIA-193, STA/EUVI-195 and STB/EUVI-195 filtergrams,
\item the coronal holes are isolated, i.e., that no other significant coronal holes were in their surrounding (Fig. \ref{37ch_1}-\ref{37ch_3}),
\item and only one distinct peak appeared in the in-situ bulk solar wind velocity measurements within 1.5 to 7 days after the center of mass of the coronal holes crossed the central meridian.
\end{itemize}
Since for each event only one significant coronal hole was at the solar central meridian and only one distinct peak appeared in the velocity measurements in the time afterwards, these coronal holes could be undoubtedly related to the in-situ measured high-speed solar wind streams.  Each of the events was re-checked manually by inspecting the solar wind bulk velocity, proton density, proton temperature, and magnetic field vector time lines and further against the ICME lists of \citet{richardson2004} and \citet{jian2013} in order to exclude ICME events; the continuously updated ICME lists can be found at \url{http://www.srl.caltech.edu/ACE/ASC/DATA/level3/icmetable2.htm} and \url{http://www-ssc.igpp.ucla.edu/~jlan/STEREO/Level3/STEREO_Level3_ICME.pdf}. Further, we re-checked each event whether the magnetic polarity of the high-speed solar wind stream matches the magnetic polarity of its source coronal hole \citep{neugebauer2002}, whereby we presumed that the magnetic polarity of a coronal hole does not change in its lifetime.
In total, this dataset covers 115 of the 594 high-speed streams measured at ACE, STEREO A, and STEREO B during the time range of interest.

\begin{figure*}[p]
\centering
\includegraphics[width = .5\textwidth]{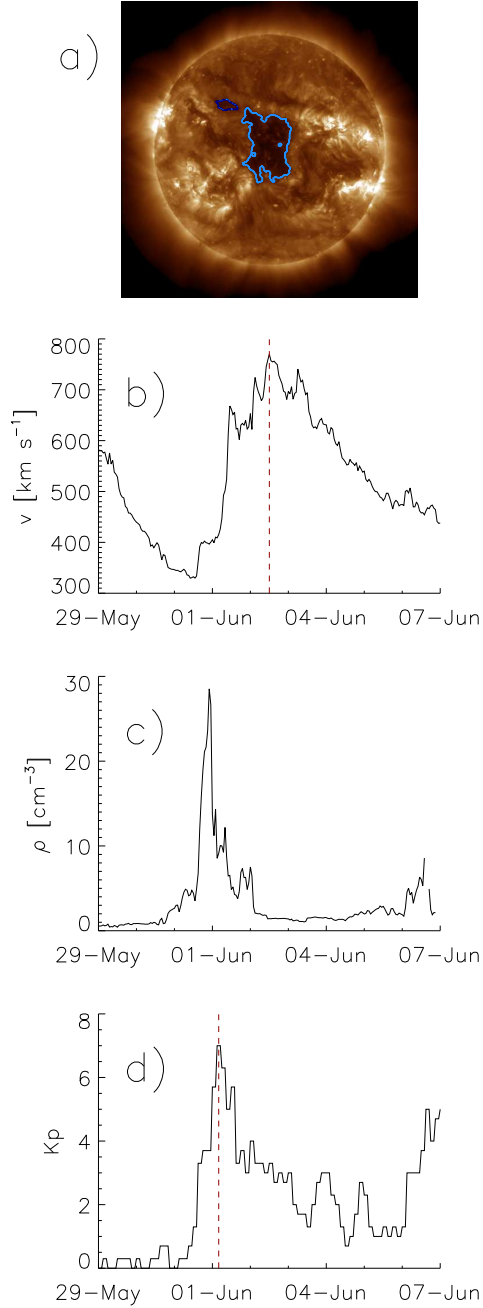}
\caption{SDO/AIA-193 filtergram of the Sun, taken on 2013/05/29. A large coronal hole is visible at the center of the disk (light blue, a). Solar wind bulk velocity measured by ACE at L1, from 2013/05/29 to 2013/06/07 (b). Solar wind bulk density measured by ACE at L1, from 2013/05/29 to 2013/06/07 (c). Kp index from 2013/05/29 to 2013/06/07 (d). The peak in the density time-line corresponds to the arriving shock front of the SIR, and the peak in the bulk velocity measurements to the arriving high-speed solar wind stream. The red dashed lines mark the peak velocity of the high-speed solar wind stream and the corresponding peak in the Kp index which are used in the analysis.}
\label{changeKp}
\end{figure*}

We extracted the borders of the 115 coronal holes under study visually by applying an intensity based thresholding technique on the AIA-193 and EUVI-195 EUV images based on \cite{rotter2012}. First, we corrected the EUV images for the EUV limb brightening due to the increased optical depth by applying the annulus limb correction \citep{verbeek2014}. Then, we extracted the coronal holes by the thresholding technique. Finally, a morphological operator with a median kernel of 9 pixels was applied \citep[Fig. \ref{changeKp}a;][]{rotter2012}. For each coronal hole, we derived its projection-corrected area $A_\text{CH}$, and the latitude of its projection-corrected center of mass $\varphi_\text{CH}$.

For each coronal hole selected, we manually assigned the peak velocity in the hourly-averaged solar wind bulk velocity data in the time-range of $1.5$ to $7$ days after the centre of mass of the coronal hole crossed the solar central meridian, and denote it as the peak velocity of the corresponding high-speed solar wind stream $v_p$ (Fig. \ref{changeKp}b). For the analysis of the strength of geomagnetic storms, we assigned the peaks of the Kp index in the time-range of $-1.5$ to $1$ days around the times of the peak velocities of the high-speed streams (Fig. \ref{changeKp}d). Note that the SIRs created by the interaction of the high-speed solar wind streams with the preceding slow solar winds are located in front of the high-speed streams (Fig. \ref{changeKp}c) \citep{belcher1971}. Therefore, the peaks in the Kp indices can already appear when the corresponding SIRs sweeps over the Earth, and thus earlier than the peaks in the velocity time lines of the high-speed streams.
 
\begin{figure*}[!p]
\centering
\includegraphics[width = .8\textwidth]{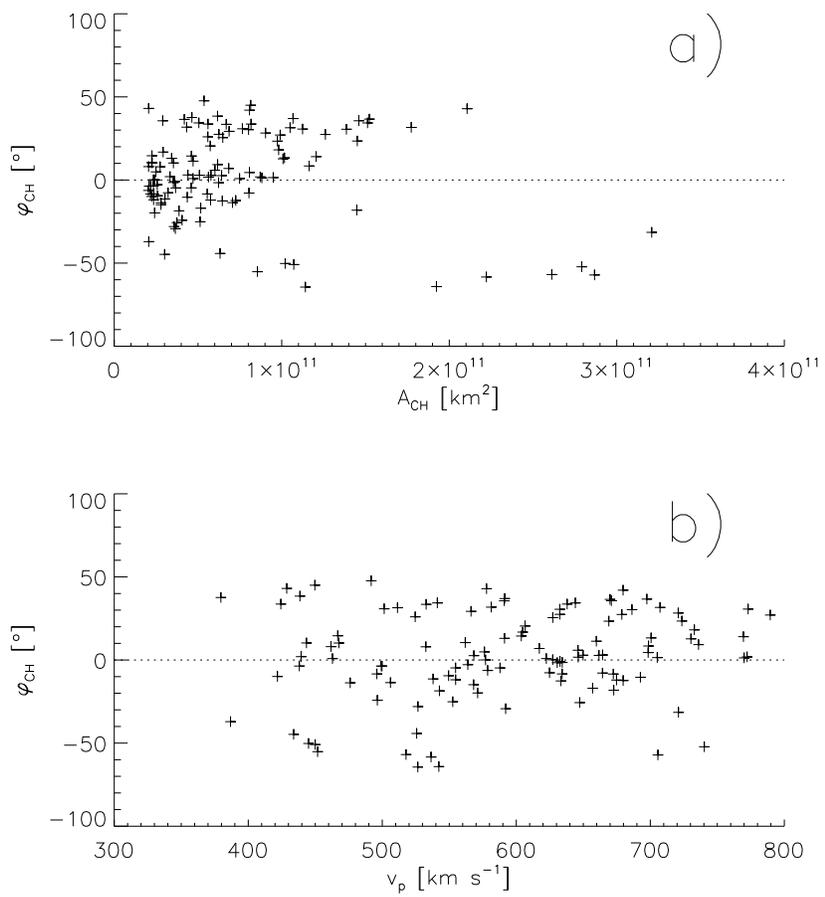}
\caption{Co-latitudes of the coronal holes versus their areas (a) and versus the peak velocities of the corresponding high-speed solar wind streams (b).}
\label{dataset}
\end{figure*}

The distribution of the solar coronal hole areas and high-speed stream peak velocities at \SI{1}{AU} versus the solar latitudes of the coronal holes are given in Figure \ref{dataset}. The EUV images of all coronal holes selected are printed in Figures \ref{37ch_1}-\ref{37ch_3}, and the properties of the coronal holes, high-speed solar wind streams and geomagnetic storms analysed are printed in Table \ref{properties_dataset}.

\section{Results}
\label{results}
\subsection{The Dependency of the High-Speed Solar Wind Stream Peak Velocities as measured in the ecliptic at \SI{1}{AU} on the Areas and Latitudes of their Solar Source Coronal Holes}
\label{hss_v}

In this section, we show that the peak velocities of the high-speed solar wind streams are dependent on the areas and the co-latitudes of their source coronal holes. We define the co-latitude of the source coronal hole as the heliospheric latitudinal angle between the position of the coronal hole and the position of the measuring satellite $\varphi_\text{co} = \varphi_\text{CH} - \varphi_\text{sat}$. Since all the measuring satellites are in the ecliptic, $\varphi_\text{sat}$ varies in the range of \SI{\approx \pm 7}{\degree}.

\begin{figure*}[!p]
\centering
\includegraphics[width = .8\textwidth]{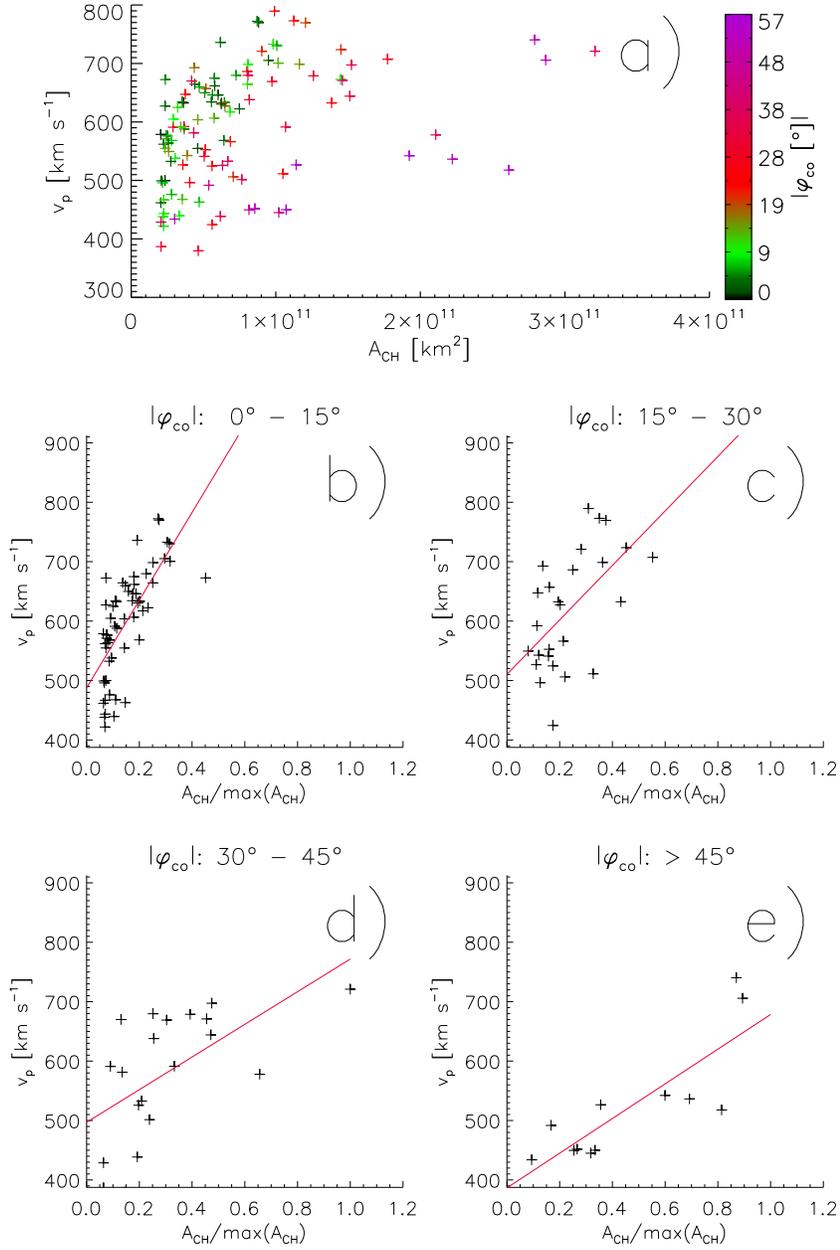}
\caption{Peak velocities of high-speed solar wind streams as measured in the ecliptic at \SI{1}{AU} versus the areas of their solar source coronal holes; the co-latitudes of the source coronal holes are coloured (a). In b-e, the dataset was divided into four panels of \SI{15}{\degree} co-latitudes, and the areas of coronal holes were normalized to the area of the largest coronal hole.}
\label{v_max}
\end{figure*}

Figure \ref{v_max}a shows the scatter plot of the peak velocities of the high-speed streams $v_p$ versus the areas of the coronal holes $A_\text{CH}$; the co-latitudes of the coronal holes are color-coded.
The well-known wide-scattered dependency between the coronal hole areas and the high-speed stream peak velocities is visible, the Spearman's correlation coefficient r$_S$ is $0.50$. However, the color coding points to a further dependence on the co-latitude of the source coronal hole.

In Figure \ref{v_max}b-e, we re-plot the coronal hole areas versus the high-speed stream peak velocities separately for coronal holes located at co-latitudes between 0$^\circ$--15$^\circ$, 15$^\circ$--30$^\circ$, 30$^\circ$--45$^\circ$, and $> 45^\circ$.
In each of the panels, the peak velocities of high-speed streams increase with increasing areas of their source coronal holes. In addition, the regression line of the $A_\text{CH}$-$v_p$ relationship is significantly steeper for coronal holes with smaller co-latitudes, i.e., for coronal holes located near the solar equator, than for coronal holes with large co-latitudes, i.e., located at medium to high latitudes. 
This means that the peak velocities of high-speed solar wind streams as observed in the ecliptic at \SI{1}{AU} do not only depend on the area, but further on the co-latitudes of their source coronal holes.  

\begin{figure*}[!p]
\centering
\includegraphics[width = .8\textwidth]{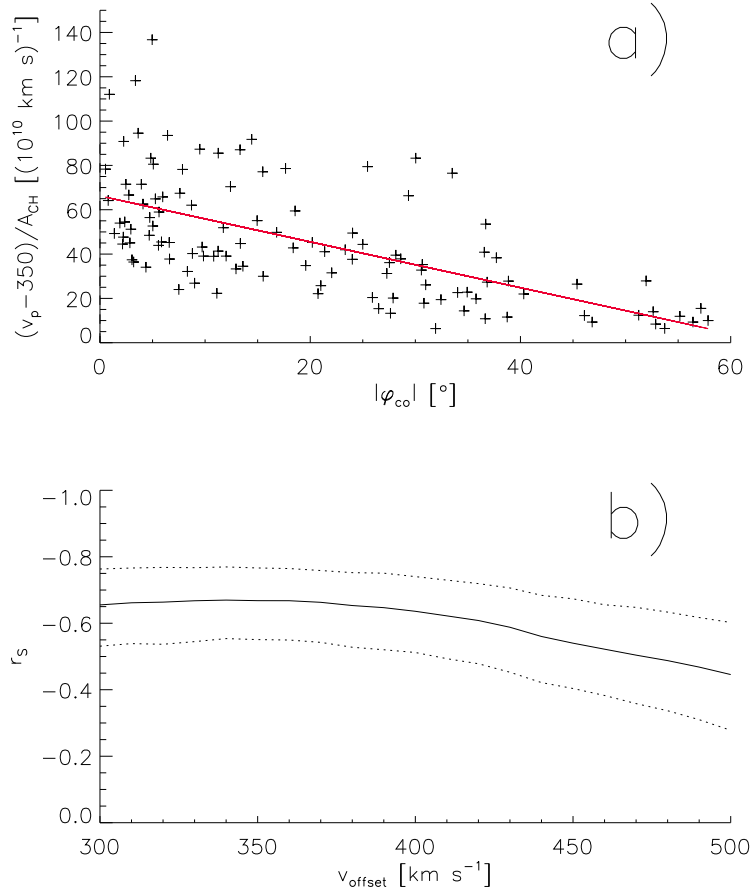}
\caption{Increase of high-speed stream peak velocity per coronal hole area as measured in the ecliptic at \SI{1}{AU}, $(v_p - v_\text{offset}) / A_\text{CH}$, versus the co-latitudes of the source coronal holes, with $v_\text{offset} = \SI{350}{km s^{-1}}$ (a). Dependence of the Spearman's correlation coefficient $r_S$ and its $0.95$ confidence interval of the dataset $\left[ (v_p - v_\text{offset}) / A_\text{CH}, |\varphi_\text{co}| \right]$ on $v_\text{offset}$ (b).}
\label{v_lat}
\end{figure*}

Next, we evaluate the relationship of the relative velocity increase per coronal hole area $(v_p - v_\textrm{offset})/A_\text{CH}$ as function of the co-latitude $\varphi_\text{co}$ of the source coronal holes. First, we presume an offset velocity $v_\text{offset}$ of \SI{350}{km s^{-1}}, and plot the relative velocity increase per area versus the absolute co-latitudes of the coronal holes (Fig. \ref{v_lat}a).
It is clearly visible that the relative velocity increase per coronal hole area depends on the co-latitude of the source coronal hole, the corresponding Spearmans's correlation coefficient is $r_S = -0.67$.
The highest relative velocity increase per area is obtained for coronal holes with small co-latitudes, i.e., located near the solar equator, and they statistically turn to zero at an absolute co-latitude of \SI{\approx 60}{\degree}.
This means that a coronal hole of a given area causes the highest high-speed stream peak velocity $v_p$ in the ecliptic at \SI{1}{AU} when it is located at the solar equator, that the peak velocity measured decreases linearly with increasing co-latitude of the source coronal hole, and that $v_p /A_\text{CH} $ even statistically turns to zero if the coronal hole is located at co-latitudes $\gtrsim \SI{60}{\degree}$.

In order to exclude that these results are dependent on the presumed offset velocity and on our manually selected dataset, we vary the presumed offset velocity $v_\text{offset}$ from \SIrange{300}{500}{km s^{-1}}, and calculate the corresponding Spearman's correlation coefficients and their $0.95$ confidence intervals by resampling the dataset \num{e5} times with bootstrapping (Fig. \ref{v_lat}b).  For offset velocities \SI{< 375}{km s^{-1}}, the Spearman's correlation coefficient stays at a high level of $\approx -0.67$ at a confidence interval of $[-0.55, -0.77]$, and decreases down to $-0.48$ at a confidence interval of $[-0.28, -0.60]$ for an offset velocity of \SI{500}{km s^{-1}}. The decrease of the correlation coefficient for high offset velocities is mainly due to the small coronal holes in the dataset: when we exclude the smallest coronal holes with $A_\text{CH} < \SI{3e10}{km^2}$, we get a correlation coefficient of $-0.61$ at a confidence interval of $[-0.46, -0.73]$ for an offset velocity of \SI{500}{km s^{-1}}. Note that for all offset velocities chosen, the relative velocity increase per area turns to zero at an absolute co-latitude of $\approx 0.60^\circ$ (not shown here). 

\begin{figure*}[!p]
\centering
\includegraphics[width = .8\textwidth]{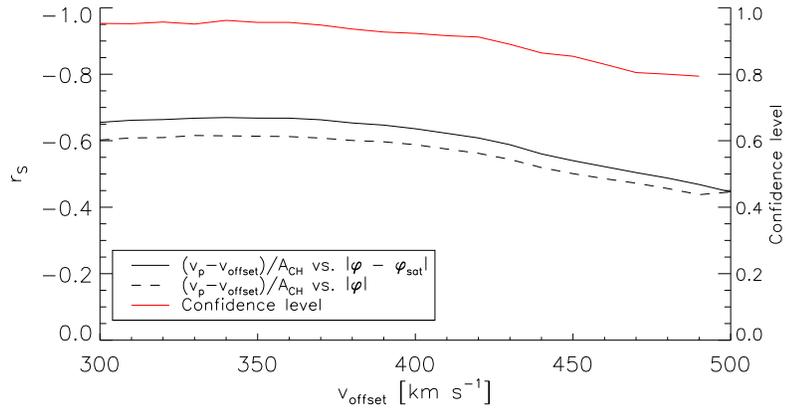}
\caption{Dependence of the Spearman's correlation coefficients $r_S$ of the datasets $\left[ (v_p - v_\text{offset}) / A_\text{CH}, |\varphi_\text{co}| \right]$ (straight black line) and  $\left[ (v_p - v_\text{offset}) / A_\text{CH}, |\varphi_\text{CH}| \right]$ (dashed black line) on $v_\text{offset}$. The red line gives the confidence level that $(v_p - v_\text{offset}) / A_\text{CH}$ is better correlated with $|\varphi_\text{co}|$ than with $|\varphi_\text{CH}|$, dependent on $v_\text{offset}$.}
\label{cc_vA_lat_satlat}
\end{figure*}

As a further test, we examine the confidence level with regard to whether the relative velocity increase per area is really dependent on the co-latitude of the source coronal hole, and not on its solar latitude. To do so, for each offset velocity, we re-sampled the dataset with bootstrapping \num{e5} times. For each sample, we calculated the Spearman's correlation coefficient between (1) the relative velocity increase per coronal hole area and the absolute co-latitude of the coronal hole, between (2) the relative velocity increase per area and the absolute solar latitude of the coronal hole, and (3) the difference of the absolute values of these two correlation coefficients. 
Then, the confidence level is given by the relative number of the samples in which the difference of the absolute correlation coefficients is positive, i.e., in which the correlation with the co-latitude is higher than with the solar latitude. The confidence level and the mean Spearman's correlation coefficients for each offset velocity are plotted in Fig. \ref{cc_vA_lat_satlat}. The red line shows that in \SI{75}{\percent} to \SI{95}{\percent} of the samples drawn, the absolute co-latitude yields a better correlation with the relative velocity increase per area than the absolute solar latitude. The black lines show that the mean difference in the Spearman's correlation coefficients is about $0.04$. Therefore, it is the co-latitude which affects the peak velocity of high-speed streams we measure in the ecliptic at \SI{1}{AU}, and not the solar latitude of the source coronal hole.

\begin{figure*}[!p]
\centering
\includegraphics[width = .8\textwidth]{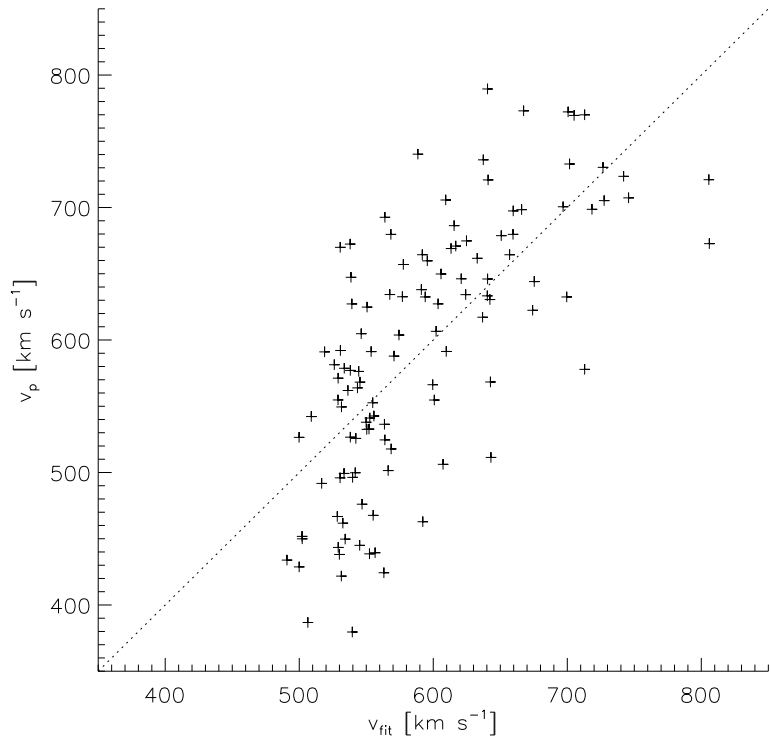}
\caption{Observed peak velocities of high-speed solar wind streams versus the peak velocities calculated by Eq. \ref{eqvmax} on basis of the source coronal hole area and co-latitude. The dashed line gives the one-to-one correspondence.}
\label{v_result}
\end{figure*}

Finally, in order to quantify the peak velocity - area - latitude dependency, we fit the data using a least-square fit and the approach $v_p = a + (b \cdot A_\text{CH}) \cdot (1 - c \cdot  \left| \varphi_\text{co} \right|) $, and get: 
\begin{equation}
v_\text{fit} [\si{km s^{-1}}]= 478 + ( \num{2.77e-9} \cdot A_\text{CH} [\text{km$^2$}]) \cdot ( 1 -  \left| \varphi_\text{co} [^\circ] \right| / \num{61.4} ). \label{eqvmax}
\end{equation}
The first term gives us the offset velocity, the second term the relative velocity increase per coronal hole area, and the third term a correction factor depending on the co-latitude of the source coronal hole. In this fit, the offset velocity should be seen as a statistical best-fit parameter without clear physical meaning. It does not correspond to the velocity of the slow solar wind, and it does not mean that high-speed streams always have a minimum peak velocity of \SI{478}{km/s}. Further, note that corresponding to the fit the relative velocity increase per coronal hole area turns to zero when the co-latitude of the source coronal hole is \SI{61.4}{\degree} with respect to the measuring satellite.
In Figure \ref{v_result}a, we show the peak velocities calculated by Equation \ref{eqvmax} versus the peak velocities observed; the dashed line marks the one-to-one correspondence. The data is well distributed around the one-to-one correspondence at a medium scatter. The Pearson's correlation coefficient of the calculated peak velocities to the measured peak velocities is $cc=0.70$, the Spearman's correlation coefficient is $r_S = 0.72$, the mean absolute error (MAE) is \SI{57}{km s^{-1}}, and the root mean square error (RMSE) of the calculated to the measured peak velocities is \SI{70}{km s^{-1}}.

\subsection{The Dependency of the Kp Index on the Area and Latitude of Coronal Holes}
\label{Kp}
In this section, we show that the strength of geomagnetic storms induced by high-speed solar wind streams are dependent on the areas and solar latitudes of the source coronal holes. Note that here we can naturally use only events recorded by SDO and ACE; thus the dataset decreases to $52$ events.

\begin{figure*}[!p]
\centering
\includegraphics[width = .8\textwidth]{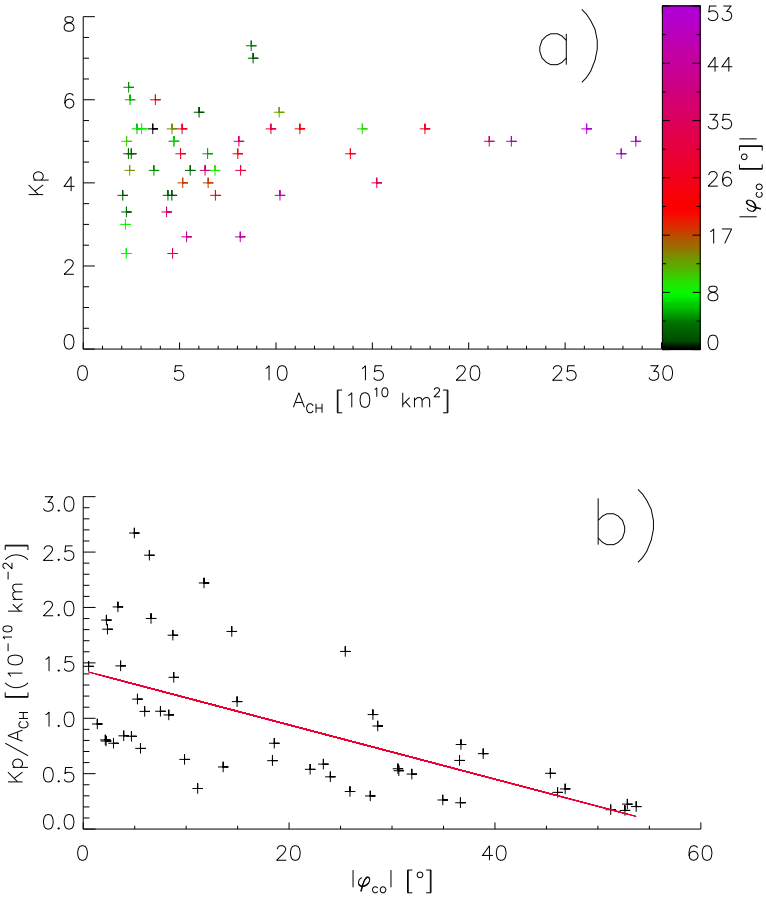}
\caption{Geomagnetic Kp index versus the areas of the source coronal holes of the corresponding high-speed streams (a).  Kp index per coronal hole area, $\text{Kp} / A_\text{CH}$, versus the co-latitudes of the source coronal holes (b).}
\label{Kp_charea}
\end{figure*}

Figure \ref{Kp_charea}a shows the Kp index versus the area of the source coronal holes of the corresponding high-speed solar wind streams, the co-latitude of the coronal holes are color-coded. A dependency of the Kp index on the area of the coronal holes is not visible.

Again, we plot the co-latitude of the source coronal holes, $\varphi_\text{co}$, versus the Kp index per coronal hole area, $\text{Kp} /A_\text{CH}$  (Fig. \ref{Kp_charea}b). Here, a dependency on the co-latitude is clearly visible at a medium scatter; the Spearman's correlation coefficient is $-0.71$ with a $0.95$ confidence interval of $[-0.55, -0.82]$. Large Kp indices per coronal hole area mainly appear when the source coronal hole is located near the solar equator, and get smaller with higher co-latitude of the source coronal hole. This means that the geomagnetic storm caused by a high-speed solar wind stream arising from a coronal hole with a given area is statistically stronger when the coronal hole is located near the solar equator than at medium latitudes, and usually weak if the coronal hole is located at higher latitudes.

\section{Discussion and Conclusions}
\label{discussion}

We investigated the dependence of the properties of solar coronal holes observed by the satellites SDO, STEREO A, and STEREO B on the peak velocity of high-speed solar wind streams as measured in-situ in the ecliptic at \SI{1}{AU} by the satellites ACE, STEREO A, and STEREO B, and the strength of their induced geomagnetic storms from 2010/08 to 2017/03. From a set of 115 solar coronal holes and corresponding high-speed solar wind streams, and from a subset of 52 geomagnetic events for the Earth-directed high-speed solar wind streams, we found that
\begin{enumerate}
\item the peak velocity of high-speed solar wind streams as measured in the ecliptic at \SI{1}{AU} depends linearly on both the co-latitude and the area of the solar source coronal hole: $$v_\text{fit} [\si{km s^{-1}}]= 478 + ( \num{2.77e-9} \cdot A_\text{CH} [\text{km$^2$}]) \cdot ( 1 -  \left| \varphi_\text{co} [^\circ] \right| / \num{61.4} ),$$
\item high-speed solar wind streams arising from solar coronal holes located near the ecliptic result in the highest solar wind peak velocities per coronal hole area in the ecliptic,
\item the high-speed stream velocity increase per coronal hole area statistically turns to zero for coronal holes located at co-latitudes $\gtrsim \SI{61.4}{\degree}$,
\item the Spearman's correlation coefficient between the high-speed stream velocity increases per coronal hole areas and the co-latitudes of the source coronal holes is higher than between the high-speed stream velocity increases per coronal hole areas and the solar latitudes of the source coronal holes, 
\item and that the Kp indices, i.e., the strength of geomagnetic storms induced by high-speed solar wind streams, depend similarly on the areas and co-latitudes of the source coronal holes. 
\end{enumerate}

Our findings are in agreement with the findings of \citet{abramenko2009}, who investigated the relation between the properties of 44 low-latitude coronal holes and the peak velocities of high-speed solar wind streams at L1. Their resulting fit gives 
\begin{equation}
v_p \text{[km/s]}= 486 + 8.5 \cdot A_\text{CH} \text{[$10^4$arcsec$^2$]},
\end{equation}
at a Pearson correlation coefficient of 0.72 between the coronal hole areas and the peak velocities.
Further, our results agree well with the results from \citet{karachik2011}, who investigated the dependency between the properties of 108 coronal holes distributed over all latitudes and the solar wind peak velocities at L1. They found that the Pearson's correlation coefficient between the area of coronal holes as measured in the image plane, $A_{ip}$, and the peak velocities is slightly higher  ($cc=0.55$) than the correlation coefficient between the projection-corrected areas and the peak velocities ($cc=0.50$). Though, they did not use explicitly a dependency on the solar latitude of the source  coronal holes. Since the coronal hole areas as measured in the image plane are related to the projection-corrected areas in the form of 
\begin{equation}
A_\text{ip} \approx A_\text{CH} \cdot \cos (\varphi_\text{CH}),
\end{equation}
their results yield implicitly a dependency of the peak velocities of high-speed streams on the solar latitude of the source coronal holes.
Our results also agree with \citet{robbins2006}, who divided the solar disk into segments of \SI{14}{\degree} longitude and \SI{30}{\degree} latitude. They calculated for each segment the fractional area covered by coronal holes, and correlated the fractional areas of the segments within meridional slices with the solar wind speeds at L1 by a multi-linear fit. They found that the weight of segments near the solar equator is higher than the weight of segments at higher latitudes, i.e., that a coronal hole with a given area results in a faster high-speed solar wind stream at L1 when the coronal hole is located near the solar equator than at higher latitudes. Due to the large size of their segments, they were not able to determine the functional relationship to the solar latitude.

Our findings confirm the well-known relationship between the areas of coronal holes and the peak velocities of high-speed solar wind streams, but additionally quantify the dependence on the co-latitudes of the measuring satellites relative to the positions of the source coronal holes on the Sun. Certainly, the solar wind speed is affected by various further parameters as the morphology of the coronal hole, which were not part of this study. 
In the following, we give an interpretation to the dependency of the peak velocity of high-speed streams as measured in the ecliptic at \SI{1}{AU} and the geomagnetic Kp index on the co-latitudes of the source coronal holes.

 The dependence on the co-latitude of the source coronal holes we found may be related to the three-dimensional propagation of high-speed solar wind streams in the heliosphere, which was studied by the Ulysses satellite. During Ulysses' fast-latitude scans, \citet{mccomas1998b} and \citet{mccomas2003b} derived the latitudinal velocity profile of two high-speed solar wind streams arising from polar coronal holes. Thereby, \citet{mccomas1998b} showed that the velocity increased sharply from the low-latitudinal slow solar wind to the polar high-speed solar wind stream, i.e., across the SIR, by $\approx \SI{190}{km s^{-1}}$ over only $\approx \SI{6}{\degree}$ latitude. However, the velocity further increased by $\approx \SI{170}{km s^{-1}}$ over a distance of $\approx \SI{30}{\degree}$ latitude inside the high-speed stream towards its center. This means that the latitudinal velocity profile in the front of the high-speed solar wind stream, i.e. the two-dimensional plane in the high-speed stream parallel to the high-speed stream - SIR interface, is not flat. Further, \citet{mccomas2003b} showed that the polar high-speed solar wind stream extended down to heliospheric latitudes of only $\approx 55-70 \si{\degree}$ strongly depending on the boundary of the polar coronal hole, i.e., it missed the Earth.

First, let us presume that in general high-speed streams propagate radially away from the Sun in three dimensions and thereby expand. In general, we expect the highest high-speed stream velocities in the center of the high-speed stream front, and lower velocities in the flanks of the high-speed stream. 
Thus, when the source coronal hole is located in the ecliptic, then the center of the corresponding high-speed stream will also propagate in the ecliptic towards our measuring satellite, and we will measure the peak velocity in the center of the high-speed stream, i.e., its maximum velocity. However, when the source coronal hole is located at medium solar latitudes, then the center of the corresponding high-speed stream will propagate radially away from the Sun towards medium heliospheric latitudes, and in the ecliptic we will only measure the flank of the high-speed stream resulting in lower peak velocities. 
The exact high-speed stream peak velocity we measure is therefore determined by the exact latitudinal position of the satellites within the high-speed stream front, which is given by the angle between the satellite and the radial propagation direction of the center of the high-speed stream. This angle equals the co-latitude we defined, i.e., the latitudinal angle between the measuring satellite and the solar source coronal hole.
Note that this interpretation is supported by the results of \citet{mccomas1998b} and \citet{mccomas2003b} described above. It is further supported by the fact that the peak velocities per coronal hole area are always correlated better with the co-latitude of the source coronal holes than with their solar latitudes, which means that the heliospheric latitudinal distance of the measuring satellite to the source coronal hole is the relevant parameter.

If our interpretation is correct, the functional dependence of the high-speed stream peak velocities on the co-latitudes of the source coronal should apply for all empirical relationships between high-speeds stream peak velocities and coronal hole parameters, in particular for relationships regarding the coronal hole area, the coronal hole brightness \citep[e.g.][]{obridko2009}, and the inverse flux tube expansion factor \citep[e.g.][]{wang1990}.

The same interpretation is valid for the dependence of the Kp index on the co-latitude and area of the coronal hole. The co-latitude determines the position of Earth in the high-speed stream front, and thus also its geomagnetic consequence.
When a coronal hole of a given area is located at the ecliptic, directly looking towards the Earth, we can expect the Earth to be directly hit by the high-speed stream with stronger geomagnetic consequences; when it is located at higher latitudes, the Earth will be farer out in the flanks of the high-speed stream and only be grazed. When a coronal hole is located at high latitudes $\gtrsim \SI{61.4}{\degree}$, the corresponding high-speed solar wind stream will eventually even not expand down to the ecliptic near Earth and thus miss the Earth.

\acknowledgments
The SDO/AIA images are available
by courtesy of NASA/SDO and the AIA, EVE, and HMI
science teams.
Full-disk EUVI images are supplied by courtesy of the STEREO Sun Earth Connection Coronal and Heliospheric Investigation (SECCHI) team.
We thank the ACE SWEPAM and the STEREO PLASTIC teams for providing the SWEPAM and PLASTIC in-situ data. 
A.V. acknowledges financial support by
Austrian Science Fund (FWF): P24092-N16. 
M.T. acknowledges the support by the FFG/ASAP Programme under grant no. 859729 (SWAMI).
S.H. acknowledges financial support by the JungforscherInnenfonds der Steierm\"{a}rkischen Sparkassen.
B.V. acknowledges financial support by Croatian Science Foundation under the project 6212 Solar and Stellar Variability.

\clearpage

\appendix
\section{Dataset}


\begin{longtable}{c r r r r r r}
  Date & Satellite &   $\varphi_\text{Satellite}$ [\si{\degree}] & $A_\text{CH}$ [\SI{e10}{km^2}] & $\varphi_\text{CH}$ [\si{\degree}] & $v_p$ [\si{km s^{-1}}] & Kp  \\
 \hline
2010-08-23 & SDO &  6 &  8.0 &  30 & 686 & 4.7 \\
2010-09-19 & SDO &  7 &  6.5 &  25 & 627 & 4.0 \\
2010-10-21 & SDO &  5 &  4.7 &  11 & 659 & 5.0 \\
2011-02-28 & SDO & -7 &  9.7 &  23 & 669 & 5.3 \\
2011-04-16 & SDO & -5 & 22.2 & -58 & 536 & 5.0 \\
2011-06-20 & SDO &  1 &  5.2 & -16 & 657 & 4.0 \\
2011-07-28 & SDO &  5 &  8.1 &  42 & 679 & 5.0 \\
2011-08-20 & SDO &  6 &  2.2 &  10 & 561 & 3.3 \\
2011-09-09 & SDO &  7 &  6.0 &   5 & 646 & 5.7 \\
2012-05-28 & SDO & -1 &  8.1 &  45 & 449 & 2.7 \\
2012-06-03 & SDO &  0 & 10.2 &  13 & 700 & 5.7 \\
2012-06-25 & SDO &  2 &  5.4 &  47 & 491 & 2.7 \\
2012-08-22 & SDO &  6 &  4.4 &   3 & 664 & 3.7 \\
2012-09-18 & SDO &  7 &  2.5 &   4 & 576 & 4.7 \\
2013-01-23 & SDO & -5 &  6.3 & -44 & 525 & 4.3 \\
2013-05-29 & SDO &  0 &  8.8 &   1 & 769 & 7.0 \\
2013-06-20 & SDO &  1 & 15.2 &  36 & 697 & 4.0 \\
2013-07-17 & SDO &  4 & 13.9 &  30 & 632 & 4.7 \\
2013-07-31 & SDO &  5 &  5.1 &  34 & 541 & 4.7 \\
2013-08-14 & SDO &  6 & 11.2 &  30 & 772 & 5.3 \\
2013-09-11 & SDO &  7 &  6.9 &  29 & 566 & 3.7 \\
2013-10-17 & SDO &  5 &  4.6 &  37 & 379 & 2.3 \\
2013-12-31 & SDO & -2 &  6.8 &   6 & 617 & 4.3 \\
2014-01-18 & SDO & -4 &  4.3 &  31 & 581 & 3.3 \\
2014-02-24 & SDO & -7 &  2.8 & -13 & 476 & 5.3 \\
2014-05-08 & SDO & -3 & 10.2 & -50 & 444 & 3.7 \\
2014-07-24 & SDO &  5 &  2.2 &  -3 & 438 & 3.0 \\
2014-11-13 & SDO &  2 &  8.2 &  33 & 637 & 4.3 \\
2014-12-04 & SDO &  0 & 27.9 & -52 & 740 & 4.7 \\
2015-01-02 & SDO & -3 & 26.1 & -56 & 517 & 5.3 \\
2015-01-30 & SDO & -5 & 28.7 & -57 & 705 & 5.0 \\
2015-02-11 & SDO & -6 &  4.7 &   0 & 462 & 5.0 \\
2015-04-17 & SDO & -5 &  2.4 & -19 & 571 & 4.3 \\
2015-05-15 & SDO & -2 &  3.0 & -11 & 537 & 5.3 \\
2015-05-24 & SDO & -1 &  2.2 &  -9 & 421 & 2.3 \\
2015-06-05 & SDO &  0 &  3.7 & -25 & 647 & 6.0 \\
2015-06-10 & SDO &  0 &  3.7 &  -4 & 587 & 4.3 \\
2015-07-02 & SDO &  3 &  5.1 & -25 & 552 & 5.3 \\
2015-07-07 & SDO &  3 &  2.3 &   0 & 627 & 4.7 \\
2015-08-12 & SDO &  6 &  2.4 &   0 & 577 & 6.0 \\
2015-10-05 & SDO &  6 &  8.7 &   1 & 772 & 7.3 \\
2015-10-10 & SDO &  6 & 21.1 &  42 & 577 & 5.0 \\
2015-10-17 & SDO &  5 &  2.1 &   8 & 461 & 3.7 \\
2015-11-07 & SDO &  3 & 17.7 &  31 & 707 & 5.3 \\
2015-11-15 & SDO &  2 &  2.3 &  14 & 466 & 5.0 \\
2015-12-12 & SDO &  0 &  4.6 &  14 & 603 & 5.3 \\
2016-05-07 & SDO & -3 &  2.4 &  -8 & 672 & 6.3 \\
2016-06-02 & SDO &  0 &  3.6 &   0 & 632 & 5.3 \\
2016-12-28 & SDO & -2 &  4.6 &  -4 & 554 & 3.7 \\
2017-01-24 & SDO & -5 &  5.6 &  -8 & 634 & 4.3 \\
2017-02-20 & SDO & -7 &  6.5 & -12 & 633 & 4.7 \\
2017-03-20 & SDO & -7 & 14.5 & -18 & 672 & 5.3 \\
2010-09-26 & STA & -1 &  5.6 &  25 & 524 & NaN \\
2010-12-03 & STA & -7 & 11.4 & -64 & 526 & NaN \\
2011-02-08 & STA & -3 & 12.6 &  27 & 678 & NaN \\
2011-04-23 & STA &  5 & 10.7 & -50 & 449 & NaN \\
2011-07-24 & STA &  4 &  8.0 &  -7 & 664 & NaN \\
2011-08-03 & STA &  2 &  4.2 &  36 & 669 & NaN \\
2011-08-26 & STA &  0 &  3.6 & -29 & 592 & NaN \\
2011-09-09 & STA & -1 &  3.9 & -18 & 542 & NaN \\
2011-09-16 & STA & -2 &  3.5 &  10 & 467 & NaN \\
2012-02-08 & STA &  0 &  2.3 &  10 & 443 & NaN \\
2012-03-31 & STA &  5 &  2.1 &  43 & 428 & NaN \\
2012-05-15 & STA &  7 &  9.8 &  18 & 732 & NaN \\
2012-06-11 & STA &  6 & 10.1 &  12 & 730 & NaN \\
2012-06-18 & STA &  5 &  2.9 &  35 & 591 & NaN \\
2012-10-14 & STA & -7 &  2.8 & -14 & 568 & NaN \\
2012-11-30 & STA & -6 &  2.1 & -37 & 386 & NaN \\
2013-02-09 & STA &  2 &  2.6 &  -2 & 563 & NaN \\
2013-04-06 & STA &  7 &  5.7 &  20 & 606 & NaN \\
2013-04-27 & STA &  7 &  6.2 &  27 & 632 & NaN \\
2013-05-13 & STA &  6 &  5.6 &   1 & 646 & NaN \\
2013-06-09 & STA &  4 &  9.5 &   1 & 705 & NaN \\
2013-07-01 & STA &  1 & 15.1 &  34 & 644 & NaN \\
2013-08-10 & STA & -3 &  6.7 &  33 & 532 & NaN \\
2013-08-25 & STA & -4 &  7.6 &  30 & 501 & NaN \\
2013-10-22 & STA & -7 &  2.0 &  -6 & 578 & NaN \\
2013-12-15 & STA & -2 &  5.1 &   2 & 649 & NaN \\
2014-01-20 & STA &  2 &  9.9 &  27 & 789 & NaN \\
2014-06-25 & STA &  0 &  8.5 & -55 & 451 & NaN \\
2014-07-17 & STA & -2 &  2.3 &  -3 & 499 & NaN \\
2015-11-29 & STA &  0 &  3.4 &  13 & 591 & NaN \\
2016-04-03 & STA &  4 & 10.5 &  31 & 511 & NaN \\
2016-07-14 & STA & -6 &  3.5 &  -1 & 634 & NaN \\
2016-08-10 & STA & -7 &  5.8 & -12 & 674 & NaN \\
2016-12-01 & STA &  3 &  3.2 &  -7 & 624 & NaN \\
2017-01-19 & STA &  7 & 32.1 & -31 & 721 & NaN \\
2017-02-09 & STA &  7 &  7.1 & -13 & 506 & NaN \\
2010-10-15 & STB &  5 &  6.2 &   9 & 736 & NaN \\
2010-12-02 & STB &  7 &  4.4 & -10 & 692 & NaN \\
2010-12-09 & STB &  7 &  2.9 &  16 & 604 & NaN \\
2011-02-21 & STB &  2 & 14.5 &  23 & 723 & NaN \\
2011-05-12 & STB & -6 & 19.2 & -64 & 542 & NaN \\
2011-07-10 & STB & -6 &  7.3 & -12 & 679 & NaN \\
2011-08-06 & STB & -4 &  6.2 &  -1 & 630 & NaN \\
2011-08-10 & STB & -4 &  3.6 & -28 & 526 & NaN \\
2011-09-01 & STB & -1 &  6.4 &   2 & 568 & NaN \\
2011-10-18 & STB &  3 &  4.0 & -24 & 496 & NaN \\
2012-02-07 & STB &  6 &  2.6 &  -9 & 549 & NaN \\
2012-02-18 & STB &  5 &  2.7 &   7 & 532 & NaN \\
2012-05-25 & STB & -5 & 12.1 &  14 & 769 & NaN \\
2012-06-21 & STB & -7 & 11.6 &   8 & 698 & NaN \\
2012-10-23 & STB &  1 &  2.4 & -11 & 554 & NaN \\
2013-01-12 & STB &  7 &  3.0 & -44 & 433 & NaN \\
2013-03-28 & STB &  3 &  6.2 &  38 & 438 & NaN \\
2013-05-19 & STB & -2 &  7.5 &   0 & 622 & NaN \\
2013-06-09 & STB & -4 & 14.6 &  35 & 670 & NaN \\
2013-06-15 & STB & -5 &  8.1 &   4 & 698 & NaN \\
2013-07-12 & STB & -6 &  3.3 &   2 & 439 & NaN \\
2013-10-27 & STB &  0 &  2.2 &  -8 & 495 & NaN \\
2013-12-01 & STB &  2 & 10.7 &  36 & 591 & NaN \\
2013-12-19 & STB &  4 &  5.8 &   2 & 661 & NaN \\
2014-01-06 & STB &  6 &  5.6 &  33 & 424 & NaN \\
2014-01-24 & STB &  6 &  9.0 &  28 & 720 & NaN \\
2014-07-20 & STB & -6 &  2.1 &  -3 & 499 & NaN \\
\caption{Properties of the coronal holes, high-speed solar wind streams and geomagnetic storms. Date denotes the date when the coronal hole crossed the central meridian, Satellite the satellite observing the coronal hole,  $\varphi_\text{Satellite}$ the heliospheric latitude of the satellite, $A_\text{CH}$ its area , $\varphi_\text{CH}$ its solar latitude, $v_p$ the peak velocity of the corresponding high-speed solar wind stream as measured in the ecliptic at \SI{1}{AU}, and Kp the geomagnetic index of the corresponding geomagnetic storm.}

 \label{properties_dataset}
\end{longtable}

\listofchanges

\end{document}